\documentclass[a4paper,11pt]{article}
\usepackage[dvipsnames]{xcolor}
\usepackage[utf8]{inputenc}
\usepackage[bbgreekl]{mathbbol}
\usepackage{geometry}
\usepackage{makecell}

\usepackage{xcolor}
\usepackage{amsmath, array, amssymb, amsfonts,amsthm}
\usepackage{upgreek}
\usepackage{xfrac}
\usepackage[inline]{enumitem}
\setlist{nolistsep}
\usepackage[french, english]{babel}
\usepackage[all]{xy}
\usepackage{txfonts}  
\usepackage{sectsty}
\usepackage{booktabs}
\usepackage{caption}
\usepackage{dsfont}
\usepackage{mathtools}
\usepackage{slashed}
\usepackage[makeroom]{cancel}
\usepackage[hidelinks]{hyperref}
\usepackage{textcomp}
\usepackage{multicol}
\usepackage[title]{appendix}
\usepackage[square,numbers,sort,compress,semicolon,merge]{natbib}
\let\cite\citep 
\usepackage{hyperref}
\hypersetup{colorlinks=true, urlcolor=blue, citecolor=blue, linktoc=page}

\usepackage{tikz}
\usepackage{tikz-cd}
\usetikzlibrary{cd}
\tikzcdset{
arrow style=tikz,
diagrams={>={Straight Barb[scale=0.8]}}
}
\usetikzlibrary{matrix,arrows,decorations.pathmorphing}

\DeclareMathAlphabet{\mathpzc}{OT1}{pzc}{m}{it}

\usepackage{fancybox}
\usepackage{mathrsfs}
\usepackage{scrextend}
\usepackage{mathrsfs}
\usepackage{footmisc}

\usepackage{mathtools}

\makeatletter
\renewcommand*\env@matrix[1][\arraystretch]{%
  \edef\arraystretch{#1}%
  \hskip -\arraycolsep
  \let\@ifnextchar\new@ifnextchar
  \array{*\c@MaxMatrixCols c}}
\makeatother

\newcommand{\defeq}{\vcentcolon=}
\newcommand{\rdefeq}{=\vcentcolon}

\newcommand\R{\mathcal{R}}
\newcommand\RR{\mathbb{R}}
\newcommand\CC{\mathbb{C}}
\newcommand\C{\mathcal{C}}

\newcommand\id{\textit{id}}
\newcommand\T{\mathcal{T}}
\newcommand\G{\mathcal{G}}
\renewcommand\H{\mathcal{H}}

\renewcommand\L{\mathcal{L}}
\renewcommand\S{\mathcal{S}}

\newcommand\SU{\mathcal{SU}}
\newcommand\U{\mathcal{U}}

\renewcommand\O{\mathcal{O}}

\newcommand\vphi{\varphi}

\renewcommand\epsilon{\varepsilon}

\newcommand\rarrow{\rightarrow}

\newcommand\aut{\mathfrak{aut}}

\newcommand\LieH{\mathfrak{h}}

\renewcommand\t{\tilde}

\renewcommand\b{\bar }

\newcommand\s{\sigma}
\newcommand\bs{\boldsymbol}

\renewcommand\-{^{-1}}
\newcommand\Ad{\text{Ad}}

\renewcommand\id{\text{id}}

\makeatletter

\newcommand{\Rmnum}[1]{\expandafter\@slowromancap\romannumeral #1@}
\makeatother

\makeatletter
\newcommand{\leqnomode}{\tagsleft@true\let\veqno\@@leqno}
\newcommand{\reqnomode}{\tagsleft@false\let\veqno\@@eqno}
\makeatother

\newcommand\diff{\mathfrak{diff}}

\DeclareMathOperator{\Diff}{Diff}
\DeclareMathOperator{\Aut}{Aut}

\theoremstyle{definition}

\makeatother

\begin{document}

\title{On the meaning of local symmetries: \\ Epistemic-ontological dialectics}
\author{J. François${\,}^{a,\,b,\,c}$, L. Ravera${\,}^{d,\,e,\,f}$}
\date{}

\maketitle
\begin{center}
\vskip -0.8cm
\noindent
${}^a$ Department of Mathematics \& Statistics, Masaryk University -- MUNI. \\
Kotlářská 267/2, Veveří, Brno, Czech Republic. \\[2mm]
${}^b$ Department of Philosophy -- University of Graz. \\
Heinrichstraße 26/5, 8010 Graz, Austria. \\[2mm]
${}^c$ Department of Physics, Mons University -- UMONS.\\
Service \emph{Physics of the Universe, Fields \& Gravitation}. \\
20 Place du Parc, 7000 Mons, Belgium. \\[2mm]
${}^d$ DISAT, Politecnico di Torino -- PoliTo. \\
Corso Duca degli Abruzzi 24, 10129 Torino, Italy. \\[2mm]
${}^e$ Istituto Nazionale di Fisica Nucleare, Section of Torino -- INFN. \\
Via P. Giuria 1, 10125 Torino, Italy. \\[2mm]
${}^f$ \emph{Grupo de Investigación en Física Teórica} -- GIFT. \\
Universidad Cat\'{o}lica De La Sant\'{i}sima Concepci\'{o}n, Concepción, Chile. 
\end{center}

\vspace{-3mm}

\begin{abstract}
 We propose our account of the meaning of local symmetries. We argue that the general covariance principle and gauge principle  both are principles of democratic epistemic access to the law of physics, leading to ontological insights about the objective nature of spacetime. 
We further argue that relationality is a core notion of general-relativistic gauge field theory, tacitly encoded by its (active) local  symmetries.
\end{abstract}

\textbf{Keywords}: Local symmetries, Bundle geometry, General relativity, Gauge field theory, Relationality.

\vspace{-3mm}

\tableofcontents

\bigskip


\section{Introduction}

While there has been a rich tradition in the philosophy of physics literature  dealing with General Relativity (GR) and its impact on spacetime ontology, see e.g. \cite{Friedman1983, Norton1987, Earman-Norton1987, Norton1988,Earman1989, Norton1993, Norton2000,  Stachel2014},
the topics of Gauge Field Theory (GFT) and gauge symmetries have been comparatively long neglected, and started to gain strong traction some 25 years ago, see e.g.~\cite{Healey1997, Lyre1999, Lyre2001, Martin2002, Castellani-Brading2003, Earman2006, Healey2009, Pitts2009, Nguyen-et-al2017}.
In~trying to extract the ontological picture of the world suggested by GR, one could count on the insights of physicists, because conceptual thinking was at the heart of the topic since its inception. 
By contrast, the interpretive efforts and attempts at conceptual clarifications of the meaning of GFT and gauge symmetries have not been made easier by the physics literature, either textbooks or specialised journals,  
where one encounters  
a fair amount of confusing and contradictory statements. Essentially, we are told that gauge symmetries are a mere unphysical redundancy of the formalism, but also that 
through the Gauge Principle (GP) 
they have been key to the discovery of the structure of matter and of the non-gravitational interactions as accounted for by the Standard Model (SM).
Such unresolved tensions persist up to this day. 
It thus falls on philosophers of physics and 
philosophically minded physicists  to try and provide a conceptually clean account of the ontology suggested by GFT. 
\mbox{As~Micheal Redhead expressed it \cite{Redhead2003}:} 
``The gauge principle is generally regarded as the most fundamental
cornerstone of modern theoretical physics. 
In~my view its elucidation is the most pressing problem in current philosophy of physics."

We propose, as physicists, what we see as the most straightforward  account of the meaning of gauge symmetries, or more generally of local symmetries, in general-relativistic gauge field theory (gRGFT), relying on the differential geometry of fiber bundles considered as its mathematical foundation.
Several authors showed early interest for the bundle geometric formulation of classical gRGFT, e.g. 
Stachel \cite{Iftime-Stachel2006, Stachel2014} for GR, and Auyang \cite{Auyang1995}, Healey \cite{Healey1997} and Lyre \cite{Lyre1999, Lyre2001}, also \cite{Catren2022}, for GFT.
We hold that the conceptual core naturally encoded in the bundle formalism is the \emph{relationality} of physics as  described by gRGFT. 
Some form of this position has been argued by others, mostly in the context of GR \cite{Rovelli1991, Rovelli2002, Earman2006, Tamborino2012}, less often in GFT \cite{Rovelli2014, Stachel2014, Rovelli2020}.
Our account emphasizes how  relationality in gRGFT arises from the dialectics between the epistemic status of local symmetries and their ontological significance.
The basic thesis we propose to defend  is the following.
Local symmetry principles are principles of democratic epistemic access to the objective geometric structure of physical spacetime.
These ``epistemic" passive local symmetries are indistinguishable from (the coordinate representation of) the local active symmetries of this structure, which, by the conjunction of what we call the  \emph{generalised hole and point-coincidence arguments}, can be naturally understood to encode the relational core of gRGFT. 

The paper is organised as follows.
In section \ref{The relational picture in general relativistic gauge field theory} we give our account of how the relational picture arises in~gRGFT. 
We first recapitulate the above mentioned dialectics in the case of general-relativistic physics in section \ref{Relationality in general relativity}.
We~use it as a template to articulate this same logic within GFT in section \ref{Relationality in gauge field theory}. There, we submit that GFT points to the geometry of an enriched spacetime as described by a principal fiber bundle. 
Finally, in section \ref{The relational core of general relativistic gauge field theory}, both pictures are brought together within the framework of gRGFT. 
There, the relational picture arising from active invariance under bundle automorphisms lead to a ``sophisticated substantivalism" towards the (principal) bundle space as describing the physical enriched spacetime, i.e. as viewing it as relationally co-defined by the  d.o.f. of physical fields. 

In section \ref{Objections}, we discuss two immediate objections to the view expressed here. 
The first, in section \ref{Associated bundles}, relates to the potential challenge presented by the alternative framing of gRGFT in terms of associated  bundles.
The~second,~in section \ref{Quantum gauge field theory}, tackles the proposition that it is  quantum gauge theories that are empirically supported, rather than classical gauge theories, so that ontological conclusions derived from the latter may be doubtful.

Finally, in section \ref{Clarifications}, after first discussing the physical relevance of Killing  symmetries in section \ref{Killing symmetries}, we~confront two topics that are testbeds for any interpretive framework for gRGFT.
In~section \ref{Spontaneous gauge symmetry breaking} we discuss   spontaneous gauge symmetry breaking, a notion which has long been considered with suspicion by philosophers of physics \cite{Earman2004a, Earman2004b, Smeenk2006, Lyre2008, Struyve2011, Friederich2013, Friederich2014, Francois2018}.
In section \ref{The Aharonov-Bohm effect}, we give a relational account of the Aharonov-Bohm effect. 
The effect is rightly considered to showcase emblematic interpretive issues of GFT and gauge symmetries, and it is the phenomenon the discussion of which has the most often mobilised the conceptual resources of  bundle geometry  \cite{Healey1997-HEANAT, Nounou2003, Healey2009, Dougherty2021}. 

In our conclusion \ref{Conclusion}, we suggest that a technical program laying ahead for whoever takes seriously the bundle-geometric-relational framework defended here has two main parts.
The first, probably easier,  is to provide a satisfying intrinsic formulation of gRGFT on a (principal) bundle space. 
The second, more challenging, is to come up with a manifestly relational framework for gRGFT, whose relationality is for now only tacitly encoded in its local symmetries. 
Such a framework would avoid misconceptions that occasionally occur in the technical literature 
and would provide a more immediate access to physical (relational) observables.


\section{The relational picture in general-relativistic gauge field theory}  
\label{The relational picture in general relativistic gauge field theory}  

Stating that Physics is relational may seem to be too obvious to stress.  
Indeed, at an elementary level, relationality in Physics may be understood to mean that:
\begin{equation*}
\label{R1}
\parbox{.85\textwidth}{{\bf Physical objects evolve \emph{with respect to each other}}.
} \tag{\text{R1}}
\end{equation*}
Yet, post special-relativistic Physics, starting with GR, adds two important refinements.  
First, it highlights that the very definition of physical objects has to be relational, so that:
\begin{equation*}
\label{R2.a}
\parbox{.85\textwidth}{{\bf Physical objects  are \emph{defined}, and evolve,  with respect to  each other}.
} \tag{\text{R2.a}}
\end{equation*}
Secondly, it  takes this refinement to its logical end point 
 in insisting that there are no physical entities that can influence others without being influenced in return; nothing can act upon without being acted upon. 
 Stated otherwise:
\begin{equation*}
\label{R2.b}
\parbox{.85\textwidth}{
{\bf There are \emph{no non-dynamical}, non-coupled physical entities}.
} \tag{\text{R2.b}}
\end{equation*}
This means that no physical structure may serve as an absolute reference, i.e. as a fixed background. 
These are furthermore unnecessary, and only arise in previous theoretical frameworks as limit cases of more fundamental dynamical entities. 

The meaning (R2) of relationality is deeper than (R1) and is often recognised as an innovative insight of GR. 
More broadly, it can be seen as a fundamental feature of the general-relativistic framework, of which GR as a theory of gravity specified by Einstein's equations, is a model. 
It is less widely appreciated that relationality (R2) is also core to gauge field theory. 
As reviewed below, in both instances relationality emerges as an ontological outcome from the requirement of 
heuristic symmetry principles that must be understood as principles of ``democratic epistemic access" to  the law of Physics, 
  according to which there are no privileged situated viewpoints on Nature. 
 The dialectics between hole arguments and point-coincidence arguments is key to the outcome.

\subsection{Relationality in general relativity} 
\label{Relationality in general relativity} 

In the interest both of a pedagogical recap and of fixing notations, let us start by reminding some elementary notions.
The~mathematical arena of GR is that of a smooth $n$-dimensional manifold $M$. 
Its local structure is given by open sets $\{U\}$ which constitute a covering of $M$. They are coordinate neighbourhoods when  endowed with a coordinate system $\{x^{\,\mu}\}: U \rarrow \RR^n$, $x \mapsto x^{\,\mu}$, and these are ``glued" together via transition functions on their overlaps $t_{ij} : U_i \cap U_j \rarrow U_i \cap U_j$. 
The automorphisms of $M$ are its group of diffeomorphisms $\Diff(M)$. 
Within a single coordinate patch, the coordinate representation of a diffeomorphism $\psi: U \rarrow U$, $x \mapsto \psi(x)$, is indistinguishable from a coordinate transformation $x^{\,\mu} \mapsto \psi(x)^{\,\mu} \rdefeq y^{\,\mu}$.

The physical content is described by fields $\upphi$ on $M$. 
Geometrically these are sections of various vector (tensor) and affine bundles: tensor fields, like the metric and electromagnetic fields,  are sections of tensor bundles, while linear connections (e.g. Levi-Civita) are sections of the affine bundle of connections. 
If the description of matter is effective, it involves vector fields (for point particles 4-velocities) or scalar fields (specifying the density of various fluids, e.g. dust, gas, etc.). A fundamental  field-theoretic description of matter involves spinor fields.\footnote{A more complete field-theoretic description of both the electromagnetic  and  matter  fields, as well as Yang-Mills fields, requires the introduction of gauge natural bundles \cite{Kolar-Michor-Slovak}, see next.}
Fields are intrinsic  objects of $M$ defined independently of any coordinate system. 
Their geometric nature  specify how their  coordinate representatives on different coordinate patches are related: e.g. tensors have  homogeneous (``tensorial") coordinate transformations.  
Some fields characterise the intrinsic geometry of $M$; the connection and the metric tensor in particular. 
To describe spacetime, $M$ is supposed equipped with a Lorentzian metric. 

Each field $\upphi$ has an orbit $\O_\upphi$, also noted $[\upphi]$, under $\Diff(M)$ s.t. for $\upphi$ a  field and $\psi \in \Diff(M)$ a diffeomorphism, $\upphi' \defeq \psi^*\upphi$ is \emph{another} tensor field on $M$. 
In general $\upphi', \upphi \in [\upphi]$ and $\upphi'\neq \upphi$.
For a given field $\upphi$, the diffeomorphisms s.t. $\upphi'=\upphi$ form a subgroup ${\sf K}_\upphi \subset \Diff(M)$, called the little group (or stability group) of $\upphi$. 
We will call it the~Killing (sub-)group. 
Its Lie algebra  $\upkappa_\upphi \subset \diff(M)$ gives the algebra of Killing vector fields of $\upphi$. 

\phantom{ciao}

Despite the fact that  in local coordinates a diffeomorphism looks like a coordinate change, they are distinct both conceptually and mathematically. 
It is clear from the fact that any intrinsic object $\upphi$ on $M$ is coordinate-invariant, but a priori transforms under $\Diff(M)$ as $\psi^*\upphi$. 
For example, a theory described by a Lagrangian density $\L(x)=\L(\upphi_{|x})$
is invariant under coordinate changes, $\L'(x')=\L(x)$, but not under $\Diff(M)$:   $\psi^*\L(x)=\L(\psi(x))\defeq \L(x')$.\footnote{In the language of differential forms, a Lagrangian $L=\L\, d^{\,n} x$ is a volume form on $M$, by definition a coordinate-invariant  object, yet subject to $\Diff(M)$, $\psi^*L \neq L$, or infinitesimally, with non-vanishing Lie derivative, $\mathfrak L_X L =d \iota_X L$ for any vector field $X\in \Gamma(TM)$. 
The~old-fashioned terminology ``passive diffeomorphisms" sometimes used to refer to coordinate changes is thus quite misleading. 
} 

\medskip
 
The above mathematical setup, supplying the kinematics of the general-relativistic framework,
 is arrived at through a well-motivated reasoning  on the general nature of the dynamics of a physical theory. 
 Let us recapitulate its logic, as it will serve as a template for a similar discussion in the context of gauge field theory.
\smallskip

Suppose that one aims to describe the dynamics of physical objects, say a set of fields, noted  $\upphi$, within a domain $U$ of spacetime with set of coordinates $\{x^{\,\mu}\}$ in terms of which one may write equations of motion (field equations) $\bs E(\upphi)=0$. 
Inputs to constrain this dynamics arise both from experimental data and from previously successful theories (expected to be recovered in some regime). Still, one might seek guiding principles to narrow down the search to a smaller space of admissible theories. 

Postulating a democracy among observers is such a general heuristic principle, 
which translates formally as the requirement
that the dynamics described should be independent of the coordinate patch one happens to be in. 
This principle of \emph{democratic epistemic access} to the laws of Physics thus amounts to requiring that the field equations $\bs E=0$ are covariant under general coordinate transformations. 
This in turn implies that the dynamical equations must be tensorial and involve objects $\upphi$ defined independently of any coordinate system, i.e. 
belonging to the intrinsic geometric structure of a  manifold $M$. 
As  coordinate-independent objects, $(M, \upphi)$ may then be interpreted as  modelling the observer-independent, objective, physical 
spacetime and its field content. 
This is a natural \emph{ontological} conclusion drawn from the \emph{epistemic} General Covariance (GC) principle:
\begin{equation*}
\label{O1}
\parbox{.85\textwidth}{
Spacetime and its field content are faithfully modelled by $(M, \upphi)$.} \tag{\text{O1}}
\end{equation*}

Yet, there is more to this story. 
As a mathematical object, $M$ has automorphisms, the diffeomorphism group $\Diff(M)$, which acts non-trivially on quantities defined intrinsically on $M$, in particular those characterising its geometry (e.g. connections and metrics): $\upphi \mapsto \psi^* \upphi$ for $\psi \in \Diff(M)$. 
It does so in a way that preserves the mathematical spaces these quantities belong to: $\Diff(M)$ preserves the spaces of connections and tensors (e.g. metrics). 
Importantly, $\Diff(M)$ preserves tensorial (field) equations, 
$\bs E(\upphi)=0 \mapsto  \psi^* \bs E(\upphi)\defeq \bs E(\psi^*\upphi)=0$. 
So, the space of solutions $\S \defeq \{ \upphi  \, |\, \bs E(\upphi)=0 \}$ is preserved by $\Diff(M)$. 
More, it is \emph{foliated into orbits}: any solution $\upphi \in \S$ has a  $\Diff(M)$-orbit $\O_\upphi \subset \S$.\footnote{The action of $\Diff(M)$ is not necessarily free: a solution may have Killing symmetries, ${\sf K}_\upphi \neq \id_M$.}
The physical ramifications of this simple fact are far reaching, 
stemming from the articulation of Einstein's famous 
 ``\emph{hole argument}" 
 and  ``\emph{point-coincidence argument}" \cite{Stachel2014, Giovanelli2021}. 

The hole argument is meant to highlight consequences of the collision between $\Diff(M)$-covariance of $\bs E(\upphi)=0$ and \eqref{O1}. 
It is usually phrased in terms of solutions $\upphi, \upphi' \in \O_\upphi$, i.e. $\upphi'=\psi^*\upphi$,  s.t.  $\psi$ is a compactly supported diffeomorphism whose support $D_\psi \subset M$ is the ``hole", which manifestly raises issues with the Cauchy problem (and in particular with the initial value problem, i.e. determinism). 
To avoid these, two options are available: 
One may {\bf (a)} renounce GC of the field equations, i.e. the principle of general epistemic democracy,  an option contemplated by Einstein between 1913 and 1915 \cite{Norton1993} (which may radically undermine \eqref{O1}). 
Or, one may {\bf (b)}  conclude that all solutions within the same $\Diff(M)$-orbit $\O_\upphi$ represent the \emph{same} physical state. 
This would imply that within a theory with generally covariant, tensorial, equations there is a one-to-many correspondence between a physical state and its mathematical descriptions. 
Such a theory, unable to physically distinguish between $\Diff(M)$-related solutions of $\bs E=0$, consequently cannot physically distinguish $\Diff(M)$-related points of $M$ either. 
In other words, $M$ is not spacetime, and one would need to update \eqref{O1} as: 
 \begin{equation*}
\label{O1b}
\parbox{.85\textwidth}{
Physical spacetime and its field content are modelled by the $\Diff(M)$-class of $(M, \upphi)$.} \tag{\text{O1b}}
\end{equation*}
 This~surjective mapping between mathematics and physics is not a vacuous ``redundancy",  as still sometimes hastily claimed, but encodes deep physical insights. 

 The final conceptual step taken by Einstein (in 1915) is what Stachel called the ``point-coincidence argument". 
 It is the seemingly obvious yet key observation that all physical interactions (hence, measurements) boil down to the spacetime coincidence of the objects (fields) involved, and that the description of these coincidences is invariant under diffeomorphisms. 
 This statement, of the $\Diff(M)$-invariance of pointwise mutual relations $\R$ between the fields in the  collection $\{\upphi\}$,\footnote{In the original formulation, the point-coincidence argument was framed as a crossing of the worldlines of  point particles: manifestly the physical event described by this meeting is $\Diff(M)$-invariant.} can be noted symbolically as
 $\R \big(\upphi; x \big) = \R \big(\psi^*\upphi; \psi\-(x)\big)$.
Or more precisely
 \begin{equation}
 \label{PC-arg}
 \begin{aligned}
 \R :  \S \times M\  &\ \rarrow \ \   \S \times M /\sim \  \ \xrightarrow{\simeq} \  \text{Relational spatiotemporal physical d.o.f.,} \\
 \big(\upphi, x \big) \  &\  \mapsto  \ \big(\upphi, x \big) \sim \big(\psi^*\upphi, \psi\-(x)\big)   \ \mapsto\   \R \big(\upphi; x \big) =\R \big(\psi^*\upphi; \psi\-(x)\big),
 \end{aligned} 
  \end{equation}
  where $\S \times M /\sim$ is the quotient of the Cartesian product $ \S \times M$ by the equivalence relation $\big(\upphi, x \big) \sim \big(\psi^*\upphi, \psi\-(x)\big)$, i.e. the space of such equivalence classes. 
  See section 3.5 of \cite{JTF-Ravera2024}, or 2.5  of \cite{Francois2023-a}, for a deeper geometrical account of such quotient spaces, understood as  ``associated bundles" to the field space bundle $\Phi$, or to the bundle of solutions~$\S$.
 The point-coincidence argument  makes {\bf (b)} the natural answer to the hole argument, 
 both dissolving  indeterminism issues and confirming \eqref{O1b}. 
 Taking this option -- preserving GC, hence $\Diff(M)$-covariance -- to its logical conclusion 
implies a two-way reading of the point-coincidence argument:  
\eqref{PC-arg}  can be understood to mean that  {\bf (i)} {physical spacetime points} (events) are \emph{individuated},  \emph{defined}, as coincidences of distinct physical (field-theoretical) d.o.f., and that {\bf (ii)} the latter are not the individual, mathematical, fields $\{\upphi\}$ but the \emph{relations} instantiated  among them. 

The ultimate ontological consequence of the epistemic GC principle, attained by the conjunction of the hole and point-coincidence arguments, is the relationality of general-relativistic physics:
\begin{equation*}
\label{O2}
\parbox{.85\textwidth}{
{\bf Spacetime is relationally defined via its field content}. \\
 {\bf Fields are relationally defined and evolve  w.r.t.  each other}. 
} \tag{\text{O2}}
\end{equation*}
Relationality is thus \emph{tacitly} encoded by the invariance, or covariance, under  $\Diff(M)$.
We may stress that relationality is obviously enjoyed in particular by solutions of $\bs E(\upphi)=0$ with Killing symmetries 
${\sf K}_\upphi \subset \Diff(M)$,
 the latter subgroup
 encoding further physical properties of the solution, as we discuss in  section \ref{Clarifications}.
The  diagram in Fig.\ref{Diag1} below summarises the logic 
leading to the relational picture. 
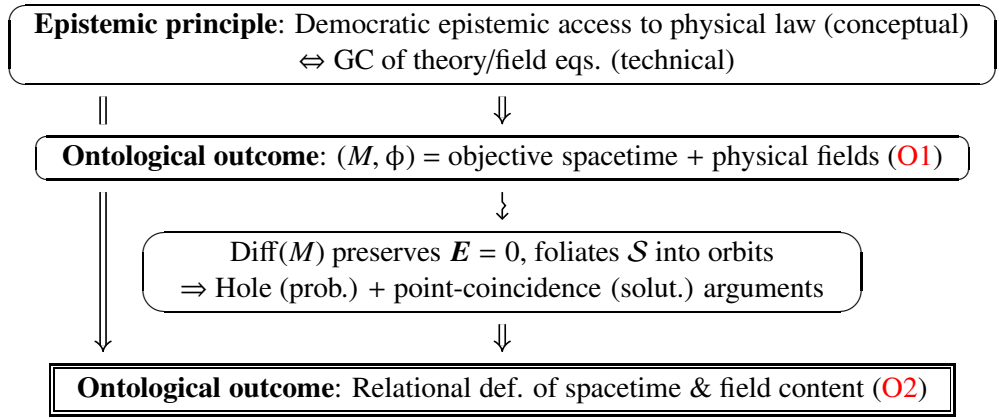
\begin{figure}[h!]
\centering
\begin{tikzcd}[row sep=1em]
\ovalbox{\parbox{\dimexpr\linewidth-20\fboxsep-200\fboxrule\relax}{\centering {\bf Epistemic principle}: Democratic epistemic access to physical law (conceptual) \\ $\quad \Leftrightarrow$ GC of theory/field eqs. (technical)}} 
\arrow[d, Rightarrow] \arrow[d, equal,  start anchor={[xshift=-5.25cm]}, end anchor={[xshift=-5.25cm]}] \\
\ovalbox{\parbox{\dimexpr\linewidth-20\fboxsep-250\fboxrule\relax}{\centering {\bf Ontological outcome}:  $(M, \upphi)$ $ = $ objective spacetime $+$ physical fields \eqref{O1}}}
\arrow[d, rightsquigarrow] \arrow[dd, Rightarrow,  start anchor={[xshift=-5.25cm]}, end anchor={[xshift=-5.25cm]}] \\
\ovalbox{\parbox{\dimexpr\linewidth-20\fboxsep-450\fboxrule\relax}{\centering  $\Diff(M)$ preserves $\bs E=0$, foliates $\S$ into orbits \\ $ \Rightarrow$ Hole (prob.) $+$ point-coincidence (solut.) arguments}}
\arrow[d, Rightarrow] \\
\doublebox{\ \text{ {\bf Ontological  outcome}: 
 Relational def. of spacetime \& field content  \eqref{O2} }\  } 
\end{tikzcd}
\caption{Relationality in the general-relativistic framework.} \label{Diag1}
\end{figure}

\noindent We now turn to the analysis of gauge theories with internal symmetries. 

\subsection{Relationality in gauge field theory} 
\label{Relationality in gauge field theory} 

Still in the interest of pedagogy, and to fix some notations, we start with a brief reminder of elementary notions about the differential geometry underlying Gauge Field Theory (GFT), before running a conceptual analysis paralleling the one of GR physics above.

The mathematical arena of GFT is that of a smooth principal bundle $P$ with structure group $H$. 
The right action of $H$ on $P$,  $R: P\times H \rarrow P$, $(p, h) \mapsto R_h p\defeq ph$, is free and transitive. 
So, it foliates $P$ into orbits, the fibers: the space of fibers, the quotient $P/H$, is a smooth manifold $M$ called the base manifold.
One thus has the projection map $\pi : P \rarrow M$, $p\mapsto \pi(p)=x$. 
Vector fields $X$ on $P$ are sections of the tangent bundle $TP$, $X \in \Gamma(TP)$. 
The tangent map of the projection is $\pi_* : TP \rarrow TM$, and its kernel define the \emph{vertical subbundle} $VP\defeq \ker \pi_*$. A~vertical vector field is a section of $VP$,  $X \in \Gamma(VP)$. 

A fiber bundle is locally trivial, i.e. over $U\subset M$ there always exist local trivialisations $t: P_{|U} \rarrow U \times H$, 
$p \mapsto t(p)= \big( \pi(p), \t t (p) \big) = \big(x, h \big)$, where $\t t:  P_{|U} \rarrow H$,
s.t. $t(ph)=t(p)h$, so $\t t(ph)=\t t(p)h$ -- meaning $\t t$ is $H$-\emph{equivariant}, $R^*_h \t t = \t t \circ R_h =\t t\, h$. 
The trivialisation $t$ provides \emph{bundle coordinates}, $\t t$ giving the fiber coordinate.\footnote{One may naturally endow $U$ with a coordinate system $\{x^{\,\mu}\}$ as in section \ref{Relationality in general relativity} above. This, together with $\t t$, supplies a full (concrete) bundle coordinate system.}
Trivialisations $t_i$ and $t_j$ over distinct overlapping opens $U_i$ and $U_j$ glue together via transition functions $h_{ij} : U_i \cap U_j \rarrow H$, $x \mapsto h_{ij}(x)$ , s.t. $t_j = t_i\, h_{ij}$ (i.e.  $\t t_j = \t t_i \,h_{ij}$). 
The set of transition functions $\{h_{ij}\}$ of a bundle reflects its topology. 
A~local trivialisation can be equivalently given by a local (trivialising) section $\s: U \rarrow P_{|U}$, required by definition to satisfy $\pi \circ \s =\id_{U}$: one then sets $\t t \circ  \s = e_H$, $x \mapsto e_H$, or $t \circ  \s = \big( \id_U, e_H\big)$, $x \mapsto (x, e_H)$. 
Two local sections $\s_i$ and $\s_j$ over $U_i$ and $U_j$ are related by a transition function $\s_j = \s_i\, h_{ij}$.
A choice of local section is thus a choice of bundle (fiber) coordinate, and changes of local sections are bundle (fiber) coordinate changes.
A choice of $t$, $\t t$, or $\s$ is sometimes  referred to as a ``choice of gauge", and bundle coordinate changes are then \emph{passive gauge transformations}. 

The maximal group of transformations of $P$ is its group of automorphisms: it is the subgroup of $H$-equivariant diffeomorphisms, preserving the fibration structure, $\Aut(P)\defeq\big\{ \psi  \in \Diff(P)\, |\, \psi(ph)=\psi(p)h\big\}$.
It thus induces~smooth diffeomorphisms of the space of fibers $P/H\simeq M$, the base manifold:
there is a surjection  $\t\pi: \Aut(P) \rarrow \Diff(M)$. 
It contains a normal subgroup, the group of vertical automorphisms $\Aut_v(P)\defeq\big\{ \psi  \in \Aut(P)\, |\, \pi \circ \psi =\pi \big\}$, noted $\Aut_v(P) \triangleleft \Aut(P)$, which induces the identity transformation $\id_M \in \Diff(M)$ on $M$: i.e. it is the kernel of $\t \pi$. Vertical automorphisms move along fibers only, they define \emph{active gauge transformations}.\footnote{One may define vertical diffeomorphisms of $P$, $\Diff_v(P)\defeq \big\{ \psi\in \Diff(P)\, |\, \pi \circ \psi =\pi \big\}$, so that $\Aut_v(P) \subset \Diff_v(P)$, which also induce $\id_M \in \Diff(M)$ on $M$. But, since these are not $H$-equivariant, they are not natural morphisms (arrows) in the category of principal bundles. 
Still, they induce \emph{generalised} gauge transformations, so are relevant for GFT. See \cite{Francois2023-b, Francois2023-a}. }
Vertical automorphisms are thus generated by the \emph{gauge group} $\H \defeq \big\{ \gamma: P \rarrow H\, |\, \gamma(ph)=h\-\gamma(p)h\big\}$ via $\psi(p)= R_{\gamma(p)}p=p \gamma(p)$. 
This is furthermore an isomorphism, $\Aut_v(P) \simeq \H$. 
We~have the short exact sequence (SES) of groups
\begin{align}
\label{SES-P}
\id_P\rarrow \Aut_v(P) \simeq \H \xrightarrow{\triangleleft} \Aut(P)  \xrightarrow{\t \pi}  \Diff(M) \rarrow  \id_M.
\end{align}
Within a 
single trivialisation, the coordinate representation of a vertical automorphism $\psi: P_{|U} \rarrow P_{|U}$, $p \mapsto \psi(p)$, is indistinguishable from a bundle coordinate transformation. 
Indeed, 
$\big(\pi(p), \t t(p)\big)= \big(x, \t t(p)\big) \mapsto \big(\pi \circ \psi(p), \t t \circ \psi(p)\big) = \big(\pi(p), \t t \circ R_{\gamma(p)} p\big) = \big(x, \t t (p) \gamma(p)\big)$. 
That is, the bundle coordinate representation of $\psi \in \Aut_v(P)$ is $\t t \mapsto \t t \gamma$, or equivalently $\s \mapsto \s \gamma$, thus resembles a  bundle coordinate transformation, i.e. a passive gauge transformation. 
\smallskip

The physical content of a GFT is described by fields $\upphi$ on $P$, which are differential forms with values in representations $(V, \rho)$ of $H$, where $\rho: H \rarrow GL(V)$ and $V$ is a vector space. 
That is,  $\upphi \in \Omega^\bullet(P)\otimes V$. 
These must be well-behaved w.r.t. the action of $H$, i.e. their equivariance is controlled by $\rho$: $R^*_h \upphi =\rho(h)\- \upphi$. 
Such forms are called \emph{equivariant}, and noted $\Omega^\bullet_\text{eq}(P, \rho)$.
The most important cases are connections $\C \subset \Omega^\bullet_\text{eq}(P, \Ad)$, and tensorial forms 
$\Omega^\bullet_\text{tens}(P, \rho)$ which are \emph{horizontal} (they vanish when evaluated on vertical vector fields $\Gamma(VP)$)\footnote{These can be seen as sections $\Gamma(E)$ of \emph{associated bundles} $E$ of $P$, a fact well-known in particular for tensorial $0$-forms (functions). See section \ref{Associated bundles} for further discussion of this point.} -- the curvature of a connection belongs to 
$\Omega^2_\text{tens}(P, \Ad)$.  
Gauge potentials $A$ and their field strengths $F$ are respectively represented by connections $\omega$ and their curvature $\Omega$, while  matter fields  and their minimal coupling to gauge potentials  are represented by tensorial forms. 
These fields are intrinsic objects of $P$, defined independently of any bundle coordinate system. 
Their geometric nature  specify how their coordinate representatives  in different trivialisations over distinct overlapping $U,U' \subset M$ are related, i.e. it determines their passive gauge transformations: 
e.g. tensorial forms have  homogeneous  passive gauge transformations (they are ``gauge tensorial", hence their name). 

Connections, and their curvatures, characterise the ``vertical" geometry of $P$.
A connection gives a prescription to identify points in neighbouring fibers (from whence their name):
two points are defined as identical if the curve in $P$ joining them is generated by a tangent vector field which is in the kernel of the connection. Such a curve is said to be \emph{horizontal} by definition. 
In other words, a connection defines a notion of horizontality in $P$. 
Following a horizontal path in $P$ circling back to a given fiber, the curvature of the connection measures the vertical distance between the initial and end points of the path. 
This is also called the \emph{holonomy} of the connection.  
In a bundle $P$ admitting a flat connection, i.e. a flat bundle, a horizontal  path circling back to a given fiber is a closed loop. 

The notion of ``sameness" across different fibers defined by a connection on $P$ (horizontality), allows therefore to define a notion of ``constancy" for any of the other $\upphi$ fields living on $P$: A connection induces a \emph{covariant derivative} preserving the mathematical spaces to which these fields belong to, $D: \Omega^\bullet_\text{eq}(P, \rho) \rarrow \Omega^{\bullet+1}_\text{tens}(P, \rho)$, $\upphi \mapsto D\upphi$. 
If~$\upphi$~ represents a matter field,  $D\upphi$ represents  its minimal coupling with the gauge potential represented by~the~connection. 
A~field $\upphi$ is said \emph{covariantly constant}, or to undergo ``horizontal" or ``parallel transport", whenever $D\upphi=0$. 
This~equation is the precise analogue of a geodesic equation within $P$.\footnote{If $\upphi$ is a spinor field, $D\upphi=0$ is essentially 
the Dirac equation (up to a mass term), the  dynamical field equation for a fermionic  field.}

Each field $\upphi$ has an orbit $\O_\upphi$, or $[\upphi]$, under $\Aut_v(P)$ s.t. for $\upphi$ 
a  field and $\psi$ a vertical automorphism, $\upphi' \defeq \psi^*\upphi$ is \emph{another} field on $P$ called the \emph{active} gauge transform of $\upphi$. 
The latter can be given explicitly in terms of the gauge group generating element $\gamma \in \H$ associated to $\psi \in \Aut_v(P)$: in particular, the $\Aut_v(P)\simeq \H$ transformations of tensorial forms are homogeneous and controlled by $\rho$ (i.e. they are ``gauge tensorial"), while that of connections are inhomogeneous (they are ``gauge pseudo-tensorial").
In general $\upphi', \upphi \in [\upphi]$ and $\upphi'\neq \upphi$.
For a given field $\upphi$, the vertical automorphisms s.t. $\upphi'=\upphi$ form a subgroup ${}^{\text{gt}}{\sf K}_\upphi \subset \Aut_v(P)\simeq \H$ that we may call Killing gauge transformations group  of $\upphi$. 
Its Lie algebra  ${}^{\text{gt}}\upkappa_\upphi \subset \aut_v(P)$ gives the algebra of Killing gauge transformations of $\upphi$.
Forms for which ${}^{\text{gt}}{\sf K}_\upphi = \Aut_v(P)\simeq \H$ are called \emph{basic}, 
noted $\Omega^\bullet_\text{basic}(P)$. 
These induces well-defined forms on the base $M$, i.e. $\Omega^\bullet_\text{basic}(P)\simeq \Omega^\bullet(M)$, meaning that they have trivial gluings or, said otherwise, they are invariant under passive gauge transformations.

Despite the fact that  in a local trivialisation a vertical automorphism, an active gauge transformation, looks like a bundle coordinate change, a passive gauge transformation, still they are conceptually and mathematically distinct.
It is obvious from the fact that any  object $\upphi$ on $P$ is invariant under change of bundle coordinates, but a priori transforms under $\Aut_v(P)$ as $\psi^*\upphi$. 
\medskip

The above mathematical setup supplies the kinematics of gauge field theory. 
The latter is arrived at via the Gauge Principle, whose logic we now recapitulate:
One aims to give a field theoretic description of experimental data, and recover (in some regime) previously successful theories. 
To do so, one posits a set of matter fields  $\upvarphi$ which represent physical plena filling a domain $U\subset M$ of spacetime  assumed to also possess non-spatiotemporal d.o.f., called ``gauge" d.o.f., which are subject to the action of a Lie group $H$ -- they thus form a representation of~$H$.\footnote{The simplest instance is that of a $\CC$-scalar field, whose phase is the gauge d.o.f., obviously subject to the action of $H=U(1)$.} 

The \emph{free} dynamics is given by field equations $\bs{\mathring E}(\upvarphi)=0$ that are $H$-covariant; $H$ thus forms a symmetry of the theory -- i.e. the Lagrangian from which $\bs{\mathring E}(\upvarphi)$ are derived is $H$-invariant.
Uneasy with the idea that the gauge d.o.f. of $\upvarphi$ would be shifted by  $h \in H$ \emph{globally}, identically, throughout the region $U$, one may suggest the gauge d.o.f. might be subject to distinct shifts from point to point, i.e. \emph{local} shifts, induced by maps $h: U \rarrow H$, $x \mapsto h(x)$. 
The infinite set of such maps forms a group under pointwise multiplication: the group of gauge transformations.
But the free field equations $\bs{\mathring E}(\upvarphi)=0$ are 
not covariant under  gauge transformations.
Covariance~is restored via the introduction of an additional (set of) field(s) $A\in \Omega^1(U)\otimes \LieH$,
minimally coupled to $\upvarphi$,  whose local gauge transformation ensures that the \emph{coupled} field
equations $\bs{E}_c(\upvarphi, A)=0$ are gauge-covariant, i.e. gauge-tensorial. 
The~field $A$ is then endowed with its own dynamics, so that the complete set of dynamical equations $\bs{E}(\upvarphi, A)=0$ is gauge-covariant.\footnote{This is done by adding the corresponding gauge-invariant term to the Lagrangian; typically the Yang-Mills term.}
These describe the dynamics and mutual interactions of $\upvarphi$ and $A$. 
By ``localising" the symmetry of a free theory, $h\rarrow h(x)$ -- ``gauging" it, in a common terminology -- one obtained an interaction theory: this is the core heuristics of the Gauge Principle (GP).

The GP is constraining, and narrows down the space of admissible theories. 
The fact that it leads to empirically adequate theories may seem like a mystery of sort, given the a priori enigmatic physical meaning of gauge symmetry, i.e. of invariance/covariance under gauge transformations.
But one may simply register this empirical adequacy of gauge theories as a raw fact and then, being informed by mathematics, attempt to understand the physical picture it outlines. 
Doing so, one soon recognises that gauge transformations are all but identical to the transition functions of a principal bundle $P$ over $M$: the field theoretic gauge transformations may be understood as  bundle coordinate changes, i.e. bundle theoretic \emph{passive} gauge transformations. 
Indeed, since the field equations \mbox{$\bs E(\upvarphi, A)=0$} are gauge-tensorial, it means that their solutions are not given uniquely, but as gauge orbits $\{\upvarphi, A\}$. 
This, one may notice, seems to lead to an issue with the well-posedness of the Cauchy problem, i.e. to indeterminism. The usual response is to accept that physics is encoded in the gauge orbit, i.e. is gauge-invariant. 
But one can indeed go a step further: as a matter of fact, the set $\{\upvarphi, A\}$  are but the bundle coordinate (local) representatives of intrinsic, bundle coordinate-invariant,  (global) objects $\upphi$  on $P$, encoding the physical d.o.f. and for whom there is no indeterminism~issue. 

Said otherwise, the space $\S_U$ of gauge field solutions on $U$ is a bundle coordinate representation of global solutions $\S$ on $P$. One may write more formally
\begin{align}
\label{loc-glob-S}
\S_U\defeq \big\{  \{\upvarphi, A\}_{|U}  \, |\, \bs E(\upvarphi, A)=0 \big\} \Leftrightarrow \S \defeq \big\{  \upphi_{|P} \rightarrow \{\upvarphi, A\}_{|U} \in \S_U \big\}.
\end{align}
One is then led to consider the bundle $P$ as modelling an \emph{enriched} spacetime, whose points are not structureless but have an internal structure (the fibers). That gauge fields $\upphi$ have ``gauge" d.o.f. is then nothing but the statement that they can probe that internal structure. 
As bundle coordinate-independent objects, $(P, \upphi)$ may then be interpreted as modelling the observer-independent, objective, internal structure of spacetime and its field content.
The GP is thus understood as a principle of democratic epistemic access to the intrinsic internal geometry of spacetime: gauge invariance of physics is but the obvious requirement of its (bundle) coordinate independence. 
To sum-up, from the epistemic Gauge Principle, and the empirical success of gauge theory, follows the ontological conclusion that
\begin{equation*}
\label{O1'}
\parbox{.85\textwidth}{$(P, \upphi)$ faithfully models the \emph{internal} geometry of an enriched spacetime and  its field content.} \tag{\text{O1'}}
\end{equation*}

Yet, there is again more to the story. 
The group of vertical automorphisms of $P$ acts non-trivially on intrinsic objects $\upphi$, preserving the mathematical spaces these quantities belong to (connections, tensorial forms, etc...). 
In~particular, since -- as reviewed above, below \eqref{SES-P} -- the bundle coordinate representation of a vertical automorphism  $\psi \in \Aut_v(P)$ is indistinguishable from a bundle coordinate transformation, 
it preserves the gauge-tensorial field equations, so its space of solutions which it \emph{foliates into orbits}:
any solution $\upphi \in \S$ has a  $\Aut_v(P)$-orbit $\O_\upphi \subset \S$.\footnote{The action of $\Aut_v(P)$ is not necessarily free: a solution may have Killing symmetries, ${}^{\text{gt}}{\sf K}_\upphi \neq \id_P$.}
The physical implications of this fact are deep and far reaching.

Analogously to GR, one may articulate an ``internal" hole argument and an ``internal" point-coincidence argument. 
The internal hole argument stresses the consequence of the collision between $\Aut_v(P)$-covariance of $\S$ and 
 \eqref{O1'}, and can  be phrased thus:
 Having two solutions $\upphi, \upphi' \in \O_\upphi$, i.e. $\upphi'=\psi^*\upphi$,  s.t.  $\psi$ is a compactly supported vertical automorphism whose support $D_\psi \subset P$ is the ``internal" hole,  manifestly raises issues with the Cauchy problem and determinism (as in the original bundle coordinate gauge orbit case). 
 Two options are again available to deal with this: 
One may {\bf (a)} drop the requirement of gauge covariance of the theory, i.e. renounce the GP  (an epistemic democracy principle), which seems incompatible with the empirical success of the gauge field theory framework. 
So, one is left with the only alternative of {\bf (b)} concluding that all solutions within the same $\Aut_v(P)$-orbit $\O_\upphi$ represent the \emph{same} physical state. 
So, in GFT,  there is a one-to-many correspondence between a physical state and its mathematical descriptions. 
Such a theory, unable to physically distinguish between $\Aut_v(P)$-related solutions, consequently cannot  distinguish $\Aut_v(P)$-related points within fibers of $P$ either. 
In other words, $P$ \emph{is not} the enriched spacetime, and one would need to update \eqref{O1'} as:
\vspace{-1mm}
 \begin{equation*}
\label{O1'b}
\parbox{.85\textwidth}{The  internal geometry of spacetime  and its fields are modelled by the $\Aut_v(P)$-class of 
$(P, \upphi)$.} \tag{\text{O1'b}}
\end{equation*}
The \emph{active} gauge covariance under the gauge group $\Aut_v(P)\simeq \H$, far from being a vacuous redundancy,  encodes a fundamental physical insight brought forth by the ``internal" point-coincidence argument.
It is this: 
Only the relative values of the internal (gauge) d.o.f. of the fields at a point $p\in P$ seem to have any physical meaning, and the description of these pointwise coincidences is invariant under vertical automorphisms.\footnote{For example, in QED, only the \emph{relative phase} of the EM potential and charged field is meaningful. See section \ref{The Aharonov-Bohm effect} for an  illustration. 
}

The statement of the $\Aut_v(P)\simeq \H$  invariance of pointwise mutual relations $\R$ between the fields in the  collection $\{\upphi\}$ we write symbolically $\R \big(\upphi; p \big) = \R \big(\psi^*\upphi; \psi\-(p)\big)$,
or more precisely
 \begin{equation}
 \begin{aligned}
 \label{int-PC-arg}
 \R :  \S \times P\  &\ \rarrow \ \   \S \times P / \sim \  \ \xrightarrow{\simeq} \  \text{Relational internal physical d.o.f.,} \\
 \big(\upphi, p \big) \  &\  \mapsto  \ \big(\upphi, p \big) \sim \big(\psi^*\upphi, \psi\-(p)\big)   \ \mapsto\   \R \big(\upphi; p \big) =\R \big(\psi^*\upphi; \psi\-(p)\big),
 \end{aligned} 
  \end{equation}
  where $\S \times P /\sim$ is the quotient of the Cartesian product $ \S \times P$ by the equivalence relation $\big(\upphi, p \big) \sim \big(\psi^*\upphi, \psi\-(p)\big)$, i.e. the space of such equivalence classes. 
 The internal point-coincidence argument makes {\bf (b)} the natural answer to the internal hole argument, 
 dissolving indeterminism issues and confirming \eqref{O1'b}.
 This option, taking the GP and $\Aut_v(P)$-covariance to their logical conclusion, implies that \eqref{int-PC-arg}  can be understood to mean that  {\bf (i)} {points of the physical internal structure of spacetime}  are \emph{individuated},  \emph{defined}, via coincidences of distinct internal physical field-theoretical d.o.f., and that {\bf (ii)} the latter do not belong to the individual mathematical fields $\{\upphi\}$ per se, but to the \emph{internal relations} instantiated among them.

The ontological consequence of the epistemic GP, reached via the conjunction of the internal hole and point-coincidence arguments, is the relationality of gauge field theoretic physics:
\begin{equation*}
\label{O2'}
\parbox{.85\textwidth}{{\bf The internal structure of spacetime is relationally defined via its field content}. \\
 {\bf Internal d.o.f. are relationally defined, and evolve w.r.t. each other}. 
} \tag{\text{O2'}}
\end{equation*}
Relationality is thus \emph{tacitly} encoded by the invariance, or covariance, under  $\Aut_v(P)\simeq \H$.
Notice that relationality is also enjoyed in particular by solutions with Killing symmetries 
${}^{\text{gt}}{\sf K}_\upphi \subset \Aut_v(P)$, which
 encode further physical properties of the solution. 
%
The  diagram in Fig.\ref{Diag2} below summarises the logic 
leading to the relational picture in gauge theory. 
\begin{figure}[h!]
\centering
\begin{tikzcd}[row sep=1em]
\ovalbox{\parbox{\dimexpr\linewidth-20\fboxsep-200\fboxrule\relax}{\centering {\bf Epistemic principle}: Democratic epistemic access to physical law (conceptual) \\  \hspace{3cm}$ \Leftrightarrow$ GP: Passive gauge invariance of theory (technical)}} 
\arrow[d, Rightarrow] \arrow[d, equal,  start anchor={[xshift=-5.75cm]}, end anchor={[xshift=-5.75cm]}] \\
\ovalbox{\parbox{\dimexpr\linewidth-20\fboxsep-170\fboxrule\relax}{\centering {\bf Ontological outcome}:  $(P, \upphi)$ $ = $ objective enriched spacetime $+$ physical fields \eqref{O1'}}}
\arrow[d, rightsquigarrow] \arrow[dd, Rightarrow,  start anchor={[xshift=-5.75cm]}, end anchor={[xshift=-5.75cm]}] \\
\ovalbox{\parbox{\dimexpr\linewidth-20\fboxsep-380\fboxrule\relax}{\centering  $\Aut_v(P)\simeq \H$ preserves  $\S$, foliates it into orbits \\ $ \Rightarrow$ Internal hole (prob.) $+$ point-coincidence (solut.) arguments}}
\arrow[d, Rightarrow] \\
\doublebox{\ \text{ {\bf Ontological  outcome}: 
 Relational def. of the internal structure of spacetime \& gauge field content   \eqref{O2'} }\  } 
\end{tikzcd}
\caption{Relationality in  gauge field theory.} \label{Diag2}
\end{figure}
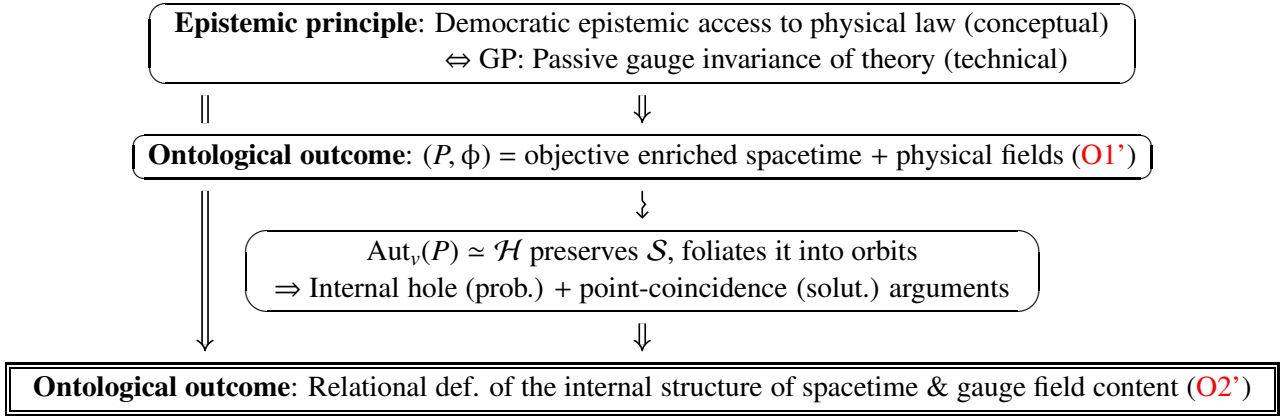

\subsection{The relational core of general-relativistic gauge field theory} 
\label{The relational core of general relativistic gauge field theory} 

\begin{figure}[h!]
\centering
\begin{tikzcd}[row sep=1em]
\ovalbox{\parbox{\dimexpr\linewidth-20\fboxsep-200\fboxrule\relax}{\centering {\bf Epistemic principle}: Democratic epistemic access to physical law (conceptual) \\  \hspace{3.5cm}$ \Leftrightarrow$ General Covariance $+$ Gauge Principle (technical)}} 
\arrow[d, Rightarrow] \arrow[d, equal,  start anchor={[xshift=-6cm]}, end anchor={[xshift=-6cm]}] \\
\ovalbox{\parbox{\dimexpr\linewidth-20\fboxsep-170\fboxrule\relax}{\centering {\bf Ontological outcome}:  $(P, \upphi)$ $ = $ objective enriched spacetime $+$ physical fields \eqref{O1"}}}
\arrow[d, rightsquigarrow] \arrow[dd, Rightarrow,  start anchor={[xshift=-6cm]}, end anchor={[xshift=-6cm]}] \\
\ovalbox{\parbox{\dimexpr\linewidth-20\fboxsep-340\fboxrule\relax}{\centering  $\Aut(P)$ preserves $\S$, foliates it into orbits \\ $ \Rightarrow$ Generalised hole (prob.) $+$ point-coincidence (solut.) arguments}}
\arrow[d, Rightarrow] \\
\doublebox{\ \text{ {\bf Ontological  outcome}: 
 Relational def. of the enriched spacetime \& gauge field content   \eqref{O2"} }\  } 
\end{tikzcd}
\caption{Relationality in  general-relativistic gauge field theory.} \label{Diag3}
\end{figure}
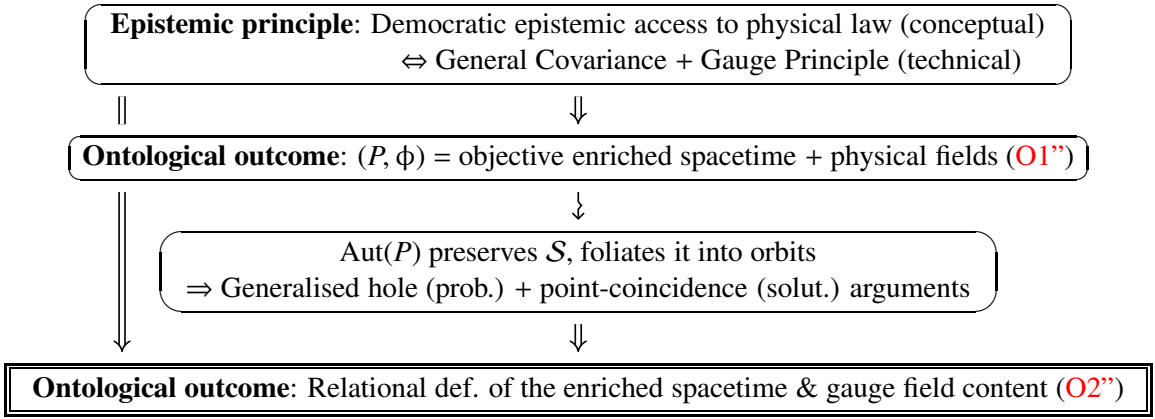

The framework of general-relativistic gauge field theory (gRGFT) is the reunion of GR and GFT. 
It can naturally be reconstructed, and/or interpreted, in a way all but similar to the above.
Starting from the principle of democratic epistemic access to the laws of physics, i.e.  GC  and the GP, one is led to the ontological picture of an objective enriched spacetime represented by a principal bundle and its field content having spatiotemporal and gauge d.o.f. characterising and/or probing its geometry. That is,
\begin{equation*}
\label{O1"}
\parbox{.85\textwidth}{$(P, \upphi)$ faithfully models the geometry of an  enriched spacetime and  its field content.} \tag{\text{O1"}}
\end{equation*}
One then appreciates that the automorphism group $\Aut(P)$ of $P$ acts non-trivially on its intrinsic objects, i.e. its field content, doing so preserving the mathematical spaces they belong to.
In particular, it preserves the solution space of the field equations of a general-relativistic gauge field theory, foliating it into $\Aut(P)$-orbits. 
From this, one formulates a \emph{generalised hole argument} establishing the tension between $\Aut(P)$-covariance and \eqref{O1"}, that is:
Having two solutions $\upphi, \upphi' \in \O_\upphi$, i.e. $\upphi'=\psi^*\upphi$,  with  $\psi$ a compactly supported  automorphism whose support $D_\psi \subset P$ is the ``bundle hole",  clearly clashes with the Cauchy problem and determinism. 
The empirical success of the theory leads to recognise that the only resolution of this tension forces to admit that physical d.o.f. have to be $\Aut(P)$-invariant, yielding the revision:
 \begin{equation*}
\label{O1"b}
\parbox{.85\textwidth}{The  geometry of physical spacetime and its fields are modelled by the $\Aut(P)$-class of 
$(P, \upphi)$.} \tag{\text{O1"b}}
\end{equation*}
Covariance under $\Aut(P)$ encodes a fundamental physical insight brought forth by a \emph{generalised point-coincidence argument}:
Only the relative values of the fields at a point $p\in P$  have a physical meaning, and the description of these pointwise coincidences is invariant under automorphisms of $P$.
The $\Aut(P)$-invariance of pointwise mutual relations $\R$ between the fields  $\{\upphi\}$ we write $\R \big(\upphi; p \big) = \R \big(\psi^*\upphi; \psi\-(p)\big)$, for any $\psi \in \Aut(P)$, or 
 \begin{equation}
 \begin{aligned}
 \label{gen-PC-arg}
 \R :  \S \times P\  &\ \rarrow \ \   \S \times P / \sim \  \ \xrightarrow{\simeq} \  \text{Relational physical d.o.f.,} \\
 \big(\upphi, p \big) \  &\  \mapsto  \ \big(\upphi, p \big) \sim \big(\psi^*\upphi, \psi\-(p)\big)   \ \mapsto\   \R \big(\upphi; p \big) =\R \big(\psi^*\upphi; \psi\-(p)\big),
 \end{aligned} 
  \end{equation}
 where $\S \times P /\sim$ is the quotient of  $ \S \times P$ by the equivalence relation $\big(\upphi, p \big) \sim \big(\psi^*\upphi, \psi\-(p)\big)$, i.e. the space of such equivalence classes.  
Taking $\Aut(P)$-covariance to its logical conclusion implies that \eqref{gen-PC-arg}  can be understood to mean that  {\bf (i)} points of the physical enriched spacetime are \emph{individuated},  \emph{defined}, via coincidences of distinct physical field-theoretical d.o.f., and that {\bf (ii)} the latter are not the individual mathematical fields $\{\upphi\}$ per se, but the \emph{relational} d.o.f. established between them.
The key ontological consequence of the epistemic general covariance and gauge principles, reached via the articulation of the generalised hole and point-coincidence arguments, is the \emph{relationality} of general-relativistic gauge physics:
\begin{equation*}
\label{O2"}
\parbox{.85\textwidth}{{\bf The physical enriched spacetime is relationally defined via its field content}. \\
 {\bf All physical d.o.f. are relationally defined, and evolve w.r.t. each other}. 
} \tag{\text{O2"}}
\end{equation*}
\medskip
The  diagram of Fig.\ref{Diag3}  summarises this logic.

The above logic, establishing relationality as the conceptual core of gRGFT, can be extended to any field theory resting on covariance or invariance under  local symmetries. 
Notably, it would apply to supersymmetric field theory and supergravity, as well as higher gauge theory,  and would  rely on supergeometry and higher geometry.

\section{Discussion}
\label{Discussion}

The picture of an enriched spacetime with structureful points -- co-defined by the gauge d.o.f. of physical fields -- emerging from gRGFT is somewhat reminiscent of the notion of extra \emph{spatial} dimensions that, after a 50 years slumber, have become quite popular again within the string theory community.\footnote{As is well-known, this idea was first introduced by Kaluza and Klein (in 1919 and 1926 respectively), at a time when GR was sparking a dual renewal: 
One in differential geometry which would ultimately lead to the theory of connections and fiber bundles, another in physics which would culminate with gauge field theory. Both strands finally reunited in the 70s. 
See the introductions of \cite{Francois2021_II} and \cite{JTF-Ravera2024review} on this point, and  e.g. the historical account (with reprints) of \cite{ORaif1997} for how the papers by Kaluza and Klein contributed to the advent of gauge theory.}
Only here, this picture emerges not from highly speculative programs, but from the well-established mathematical framework nesting our best (most accurate) 
 theories of fundamental physics.

One may furthermore observe that the kind of internal structure for spacetime points hinted at by gRGFT is more subtle than mere additional spatial dimensions. 
If tentative  programs like string theories were, in the fullness of time, proven correct, this would arguably be deflationary of the radical novelty of the spacetime picture brought by gRGFT.  
As of now, enthusiasm for extra-dimensions has at least be serviceable in possibly removing from the mind of many the psychological impediment to accepting the  enriched spacetime of gRGFT. 
But there maybe better objections, the two most immediate of which we want to address next, before turning to further clarifications and observations entailed by this view. 

\subsection{Objections}
\label{Objections}

\subsubsection{Associated bundles}
\label{Associated bundles}
In section \ref{Relationality in gauge field theory}, we stated that the physical gauge fields are described by fields $\upphi$ on the principal bundle $P$ which are equivariant forms $\Omega^\bullet_\text{eq}(P, \rho)$, with matter fields and their minimal coupling in particular represented by tensorial forms $\Omega^\bullet_\text{tens}(P, \rho)$. 
We noted that the latter can also be seen as sections of vector bundles said to be \emph{associated} bundles of $P$. 
Let us remind how they are defined. 

Given a representation $(\rho, V)$ of the structure group $H$ of $P$, one defines on the direct product $P\times V$ a right action of $H$ as: $R_h\,(p, v)\defeq \big(ph, \rho(h)\-v\big)$, $ \forall h\in H$.
This action of $H$ defines an equivalence relation on $P\times V$: $\big(ph, \rho(h)\-v\big)\sim (p, v)$. 
The  bundle over $M$ associated to $P$ via $(\rho, V)$ is defined as the vector bundle given by the quotient of the product space by the above equivalence relation:  $E\defeq P\times V\,/\!\sim$, also often noted  $E=P\times_\rho V$. 
A~point of $E$ is an equivalence class $[p, v]$ under $H$, and its projection map is defined as $\t\pi ([p, v])\defeq \pi(p)=x \in M$. 
A fiber over $x\in M$ is defined as $E_{|x}\defeq  {\t \pi}\- (x)$, and is non-canonically isomorphic to $V$ 
(thus called the ``typical fiber" of $E$).

The~linear space of sections of $E$ is $\Gamma(E)\defeq \big\{ s: M \rarrow E\, \ |\ \t \pi \circ s =\id_M \big\}$. 
It is  an elementary result of bundle theory that there is an isomorphism $\Gamma(E)\simeq \Omega^0_\text{tens}(P, \rho)$: Indeed for $\upphi \in\Omega^0_\text{tens}(P, \rho)$, i.e. satisfying $\upphi(ph)=\rho(h)\-\upphi(p)$, or $R^*_h\upphi =\rho(h)\-\upphi$, on $P$, one may form the section $s(x)\defeq \big[p, \upphi(p)\big]=[ph,\upphi(ph)]\, \in \Gamma(E)$. The inverse if clear.  
Matter fields are thus represented also by sections $\Gamma(E)$ of associated bundles $E$.

A~connection on $E$ is defined as the linear operator on sections 
$\nabla: \Gamma(TM) \times \Gamma(E) \rarrow \Gamma(E)$, $(X, s)\mapsto \nabla_X s$,
satisfying $C^\infty(M)$-linearity in the first argument and a Leibniz-type relation in  the second: 
$\nabla_X (fs)=X(f)s + f\nabla_X s$, $\forall f \in C^\infty(M)$. The object $\nabla s \in \Omega^1(M)\otimes E$ represents  the minimal coupling of matter fields with a gauge potential encoded in $\nabla$. Its curvature is $F \defeq \nabla \circ \nabla \in\Omega^2(M)\otimes \text{End}(E)$, i.e. $\nabla^2s= Fs$. 
A section is said to be parallel transported, or constant, if $\nabla s=0$: i.e. a connection $\nabla$ defines a notion of ``sameness" across different fibers in~$E$. 
One shows that there is a 1:1 mapping between principal connections on $P$ and connections $\nabla$ on its associated bundles $E$. Actually a connection on $P$ will induce connections on all its associated bundles. All structures present on $P$ are in 1:1 correspondence with structures present or induced on $E$.

For example, the tangent bundle $TM$ and cotangent bundle $T^*M$ are associated bundles to the frame bundle $P=LM$, with structure group $H=GL(n)$, via the fundamental (dual) representations $V=\RR^n$ and $V'=\RR^{n*}$. 
Any tensor bundle $\T M$ over $M$ is likewise associated to the frame bundle. 
Spinor bundles $\sf S$ are associated to the orthonormal frame bundles $OM$ with structure group $S\!O(r, s)$ -- a reduction of $LM$ -- or to its cover, the spin bundle (or \emph{spin structure}) ${\sf S}M$, via spin representations $(\rho, S)$ of  $S\!O(r, s)$ or its spin group Spin$(r,s)$. 
Spinor fields can thus be described either as sections of spin bundles $\Gamma(\sf S)$,
or as tensorial 0-forms on $OM$ (or ${\sf S}M$).

For realistic theories, the principal bundle one considers is $P=OM\times Q$, with structure group $H=S\!O(1,3) \times G$, where $G$ is an ``internal" group -- in the Standard Model $G=U(1) \times S\!U(2) \times S\!U(3)$. 
Matter fields are equivalently represented by tensorial 0-forms $\Omega^0(P, \rho)$ for $H$-representations $(\rho, S\otimes V)$,  or by sections $\Gamma(E)$ of the associated bundle $E=P\times_\rho (S\otimes V)$. 
Gauge potentials are represented either by connections on $P$ or on $E$. 
 \medskip

One may wonder if the fact that there are two mathematically equivalent ways to formulate gRGFT -- either in terms of a principal bundle $P$, or in terms of an associated bundle $E$ -- undermines the ontological commitment to the formalism as derived in the previous section.
Naturally, it depends on which part of the formalism. 
One may commit to what is common to
both equivalent formulations: 
the fact that spacetime is enriched with structureful points, and that gauge d.o.f. of fields  relationally co-define (or probe) this internal structure. 
The choice $P$ \emph{vs} $E$ would be within the reasonable, and maybe incompressible, bounds of conventional choice of mathematical representation of the ontological ground truth  revealed by the empirical success of gRGFT (as we understand it).  

The arbitration between the two may for now rely on aesthetic and pragmatic considerations. 
In our view, the principal bundle $P$ retains its primacy as the most natural and economical representation of the physics of gRGFT. 
Indeed, arguably a principal bundle $P$ has conceptual priority over, and controls, its various associated bundles -- in the same way that a group has logical priority over  its representations. 
Furthermore, the characteristics of the \emph{internal} structure of spacetime are more directly read from $P$ than from $E$: notably its  dimensionality. This is due to the fact the action of $H$ on fibers of $P$ is always \emph{free}, but often only \emph{transitive} on its orbit in fibers of $E$ (e.g. in QCD).\footnote{A well-known nice feature of YM-type gauge theories is the heuristics according to which the dimension of the  group $H$ counts the number of interaction mediating gauge fields.
Actually, one may say that it is rather the dimensionality of the fibers of $P$ doing this counting. 
}
Also, the full gauge structure of gRGFT emerges more immediately and naturally from the geometry of~$P$. 


\subsubsection{Quantum gauge field theory} 
\label{Quantum gauge field theory} 
An obvious caveat that may come to mind, regarding the idea of taking seriously the bundle geometry as we argue above, is the following.
When saying that gRGFT has  strong empirical suppport, we actually mean that two theoretical models within the framework are incredibly accurate: GR and the SM. 
Yet, the empirical accuracy of the SM stems from it being a 
\emph{quantum} gauge field theory. 
The bundle geometry is a feature of its classical limit (its Lagrangian), so maybe it should not be taken all that seriously.
We see a number of immediate obvious  answers. 

First, as QFT remains an unfinished unification of Quantum Mechanics and Special Relativity (SR), without a firm and definitive mathematical foundation, one must still rely on heuristic quantization procedures that take a classical theory (a Lagrangian) as input, and output a QFT (typically a path integral, Feynman graphs, etc.). 
What~these procedures do, is to place quantum constraints on the behavior of classical d.o.f., altering aspects of their nature, or properties, but they do not radically undermine their existence.
The SM, as a QFT, thus relies in an essential way on  geometric ingredients integral to the ontological picture depicted above: gauge d.o.f. and gauge symmetries. 
One is hard pressed to see how the success of the SM is any less than a quite direct support for the ontological picture of the key classical features on which it rests. 

Futhermore, there are ``qualitative" features of the SM, that stems from the classical theory; so-called ``tree-level" features, or predictions. 
Notably, the prediction of the very existence of some fields (particles); 
either as part of some gauge multiplets (e.g. charm or top quarks, Higgs), 
or as part of interaction mediating gauge fields (the $Z^0$ as the carrier of neutral currents). 
One may also think e.g. of the mass ratio of $W^\pm$ and $Z^0$ bosons (Weinberg angle) in the electroweak model. 
Another such qualitative feature is the  ``flavour universality" of couplings between matter and gauge potentials (lepton and quark universality), stemming from the description of these (minimal) couplings via covariant derivatives. 
Also, the allowed branching decays for any given particle/field are given by the classical theory, because it specifies the allowed  interactions (even though the precise branching \emph{ratios} need QFT computations).
These (verified) tree-level features and predictions hint at the fact that the qualitative picture -- the number, nature, organisation and possible interactions of the internal/gauge d.o.f. -- stemming from the classical/geometric description is, for all we currently know, correct. 
This again lands some support to the ontological view derived from the bundle geometric framework. 

One may also highlight that there is actually a part of the SM that has a well-defined classical limit describing a long range interaction: the quantum electrodynamics sector (QED). 
Its classical limit is Maxwell's EM theory, whose underlying geometry is that of a $U(1)$-principal bundle on which  charged Dirac spinors (representing matter fields) live. Classical EM by itself already has  significant empirical support.
If anything, this may speak quite directly in support of the view of an enriched spacetime whose internal structure is that of a compact 1-dimensional space, and whose geometry is responsible for the long range EM interactions. An idea not far from the original visions of  early pioneers such as Weyl, Kaluza and Klein \cite{ORaif1997}.

Finally, most believe that there is a quantum theory of gravitation, quantum gravity, whose limit is GR as described by its natural geometric formulation. 
Yet, most of the same are not tempted to reject the adoption of an ontological picture of spacetime as understood through  the geometric formulation of GR, on the ground that a more fundamental quantum theory (probably) exists. 
One is usually happy to accept that fundamental quantum d.o.f. give rise in the limit -- in some emergent way perhaps --  to a  classical field/spacetime, with its geometric qualities. Such is incontrovertibly the case for QED and classical electrodynamics. 
In the case of the SM as a gauge QFT,
even though the two nuclear interactions have no long range classical manifestation, the ontology of their classical/geometric limit may be taken seriously, as a key component of the explanation of their empirical success. 

\subsection{Clarifications}
\label{Clarifications}

Admitting that the above is enough to assuage a first wave of skepticism 
towards the logic of Fig.\ref{Diag3} and the conclusion \eqref{O2"}, let us go on and discuss 
within this interpretive framework two topics that have often been considered especially relevant to the philosophy of gRGFT:  
spontaneous gauge symmetry breaking (SSB) and the Aharonov-Bohm (AB) effect. 
But first, we may consider the less widely appreciated issue  of Killing symmetries.

\subsubsection{Killing symmetries}
\label{Killing symmetries}
 One may perceive some tension between claiming that physical d.o.f. must be  $\Diff(M)$- and $\Aut_v(P)$-invariant, and the claim that the Killing symmetries of a field configuration $\upphi$, 
${\sf K}_\upphi \subset \Diff(M)$
or 
${}^{\text{gt}}{\sf K}_\upphi \subset \Aut_v(P)$, are physically meaningful. 
If the physical d.o.f. encoded in, or represented by, $\upphi$ are all invariant, why would Killing symmetries matter at all? 
The tension is only superficial, stemming from hastily thinking of $\Aut_v(P)$ and/or $\Diff(M)$ as ``non-physical" redundancies. 
%
As we have argued, the $\Diff(M)$/$\Aut_v(P)$-covariance of the field equations $\bs E(\upphi)=0$ of a gRGFT encodes the relational character of the fundamental physical d.o.f. of their solutions. This relationality is enjoyed  in particular by 
solutions with Killing symmetries, but the Killing subgroups ${\sf K}_\upphi/{}^{\text{gt}}{\sf K}_\upphi$ encode further physical properties of such solutions: notably they 
signal that $\upphi$ may have a privileged class of ``observers". 
So, both $\Diff(M)$/$\Aut_v(P)$ and ${\sf K}_\upphi/{}^{\text{gt}}{\sf K}_\upphi$ encode physical information, as there are of course relevant physical differences between solutions $\upphi$ with distinct Killing groups.

Let us consider e.g. the general relativistic framework, in the standard metric formulation; $\upphi=\{g, \ldots \}$, with the fields besides the  metric $g$ left unspecified. 
The $\Diff(M)$-covariance of the field equations $\bs E(g, \ldots)=0$ of a given theory implies the relational co-definition of the d.o.f. encoded in $\upphi=\{g, \ldots \}$. 
This holds true in particular for metric solutions with non-trivial Killing groups ${\sf K}_g \neq \id_M$. 
But on top of this, 
the Killing group of a metric can signal (or designate, or select)  classes of special ``observers" for whom the geometry, i.e.  $g$, is simplest:
By~which we mean that that in the coordinate system derived from a given (set of) Killing diffeomorphism(s), the metric has the least possible number $\#$ of non-zero components --  $\#\in \{n , \tfrac{n(n+1)}{2}\}$ -- and these are functions of the least possible number of coordinates.\footnote{Of course, in practice, to find exact solutions of the field equations, typically the logic is reversed: One chooses a priori a coordinate system, and imposes symmetry considerations lowering $\#$. Then one requires further conditions (e.g. stationarity) implying that the holonomic vector fields associated to some of the coordinates are Killing vector fields of the sought after metric.
}
If one of those Killing diffeomorphisms is  time-like, generated by $\xi  \in \upkappa_\upphi \subset \diff(M)$ s.t. $g(\xi, \xi)\!>\! 0$, it represents a potential physical observer for whom the geometry of spacetime looks especially simple -- it is an ``adapted observer". 
Killing diffeomorphisms also signal the existence of covariantly conserved quantities as measured by observers in geodesic motion w.r.t. to the Levi-Civita connection $\nabla$ induced by $g$: 
Indeed, for  $X$ generating a geodesic, $\nabla_X X=0$, the quantity $q\defeq g(X, \xi)$ is conserved. If $X$ is time-like, representing a potential physical observer, the latter will measure $\nabla_X q=X(q)=0$.
A related result is that any Killing vector field with constant length, $X\big( g(\xi, \xi)\big)=0$ $\,\forall X$, generates a geodesic flow, i.e. $\nabla_\xi \xi=0$.
If such a $\xi$ is time-like, it can represent a physical observer in natural motion in the geometry and for whom it again appears ``simple". 

In GR, with $\bs E(g, \ldots)=0$ the Einstein field equations, such solutions with ${\sf K}_g \neq \id_M$ are  actually  all known exact solutions of the theory (e.g. Schwarzschild, Kerr, deSitter, FLRW cosmological solutions, etc.). 
All models are nonetheless relational in an essential way. 
SR can be seen as a model of the general relativistic framework, with field equation for the metric $\bs E(g)=\text{Riem}(g)=0$.
As such, as stressed by a ``Kretschmann objection", SR in fact enjoys $\Diff(M)$-covariance, thus is relational.
What makes it special is that: 
1) The field equation decouples the  dynamics  of metric from  that of other fields (they do not act as source for the metric). 
2) The (unique up to $\Diff(M)$) solution of the metric field equation, the Minkowski metric $g =\eta$, has a  frozen dynamics (no d.o.f.) because its Killing group, the Poincaré group ${\sf K}_\eta = IS\!O(1,n\!-\!1)$, has maximal dimension.
Yet,  the remaining dynamical objects are coupled to the metric still:  via the geodesic equation (minimal coupling) for point particles, or, for fields, through the appearance of $\eta$ in their field equations. 
The metric field $g=\eta$  thus becomes a background structure: it acts upon, still dictating the causal and inertial structure, without being acted upon. 
Physical spacetime points remain relationally defined via field coincidences, only this time the homogeneity and lack of dynamics of $\eta$ make it effectively unsuited to participate efficiently in their labelling, so one must rely on the remaining dynamical objects to do so. 

\subsubsection{Spontaneous gauge symmetry breaking}
\label{Spontaneous gauge symmetry breaking}

The notion of gauge SSB is considered by many a central notion of the SM or particle physics. This is due to it being understood as a key component of the BEHGHK mechanism of mass generation in the electroweak (EW) sector:  
A change in the shape of the potential of the scalar (Higgs) field -- an ``EW vacuum transition" -- endows it with a non-zero vacuum expectation value (VEV), which simulatenously 1)
breaks the gauge group $\SU(2) \times \U(1)$ of the model down to the electromagnetic $\U(1)$ subgroup, and 2) endows all fields coupled with the scalar field with  masses proportional to its VEV and to the constants measuring the strength of their couplings.\footnote{For almost all elementary particles, the notion of mass is thus replaced by that of coupling to the scalar field. The exception is neutrinos; the origin of their masses remains an open problem.}  
The predictions of the EW model have been continuously validated, most notably in 1973 with the discovery of the neutral current, in 1983 with the detections of the weak bosons $W^\pm$ and $Z^0$, and finally in 2012 with the detection of the Higgs boson, the quantum of the scalar field. 
Not mentioning the manifold  high precision measurements derived from the model. 

This seems to land strong support  to the SSB interpretation of the EW model. 
Yet, there is a curious tension accross the physic literature, as on the one hand we are often  told that gauge symmetries are not true symmetries of Nature but a mere redundancy of the formalism, 
but on the other hand SSB is as often celebrated as a key physical mechanism with unquestionable empirical support. 
Philosophers of physics were quick to notice. Earmann is often cited as spearheading the call for resolving this inconsistency, twenty years ago: 
``[...] a genuine property like mass cannot be gained by eating descriptive fluff,
which is just what gauge is. Philosophers of science should be asking the Nozick question: What is the objective
(i.e., gauge invariant) structure of the world corresponding to the gauge theory presented in the Higgs mechanism?" \cite{Earman2004b}.
He also suggested that the idea of gauge SSB may not survive a closer analysis, laying arguments that, 
in his view,
``[...] suffice to plant the suspicion that when the veil of gauge is lifted, what is revealed is that the Higgs
mechanism has worked its magic of suppressing zero mass modes and giving particles their masses
by quashing spontaneous symmetry breaking. However, confirming the suspicion or putting it to rest
require detailed calculations, not philosophizing."
\cite{Earman2004a}.

In the following years, several authors took up the challenge \cite{Smeenk2006,Lyre2008, Struyve2011, Friederich2013, Friederich2014}, showing indeed that one can formulate the EW model in an $\SU(2)$-invariant way, decoupling the notion of $\SU(2)$ breaking, or reduction, from that of EW vacuum transition. The latter being shown to be the operative mass generating mechanism.  
This insight actually laid dormant in neglected corners of the physics literature: Higgs \cite{Higgs66} and Kibble \cite{Kibble67} already stressed the fact even before the EW model was proposed. 
In the following decades, many rediscovered the fact or hinted at it \cite{Banks-Rabinovici1979, Frohlich-Morchio-Strocchi80, Frohlich-Morchio-Strocchi81, GrosseKnetter-Kogerler1993, Buchmuller-1994, McMullan-Lavelle95, Chernodub2008, Faddeev2009, Masson-Wallet}. 
See \cite{Francois2018, Berghofer-et-al2023} for details. 
\medskip

How does all this fit the ontological geometric picture of an enriched spacetime as defended here?
On this view, gauge symmetries are not mere ``redundancy" or ``descriptive fluff', they encode the relationality of gauge physics, and point to the internal structure of spacetime, as described by bundle geometry.
The internal structure of spacetime points hinted at by the gauge group of the SM is that of a compact space $S^1 \times S^3 \times \{{\sf P}^8 \xrightarrow{\text{\tiny $S^3$}} S^5\}$ diffeomorphic to $G=U(1)\times S\!U(2) \times S\!U(3)$, 
where the compact connected 8-dimensional space $\sf{P}^8$ is a non-trivial bundle over $S^5$ with fibers $S^3$ -- i.e. a non-trivial $S\!U(2)$-principal bundle over $S^5$.\footnote{Possibly, the real internal space is a projectivisation of the above, if one insists that the true gauge group of the standard model is $\G/N$, with $N \subset Z(\G)$ a discrete subgroup of the center  $Z(\G)$ of $\G=\U(1)\times \SU(2) \times \SU(3)$. See e.g. \cite{Baez2005, Tong2017}.}

In this picture, there is a priori less of a tension with the idea of gauge SSB. 
Taken  uncritically -- i.e. admitting for the sake of discussion that there is indeed a dynamical mechanism by which the $\SU(2)$-gauge 
symmetry is broken in the EW model -- a possible interpretation would be that what this mechanism actually reflects is a dynamical change in the internal geometry of spacetime. In regions of spacetime where the VEV transitions from zero to non-zero, there would be a  collapse of the 3-sphere $S^3$ part of spacetime fibers (which is diffeomorphic to $S\!U(2)$). The SSB mechanism would actually reflect a true phase transition of the internal geometry of spacetime, from a state in which it is described by a $G$-bundle to a state in which it is described by  a $G/S\!U(2)$-subbundle.
However, on the idea that gauge d.o.f. of fields and their interactions relationally co-define (and probe) the internal structure of spacetime, one may worry that this interpretation seems to entail a ``switching off" of weak interactions as the $S^3$ fibers collapse, which is obviously not what is observed, on the contrary.

A refined position is allowed if one takes a more critical stance on the idea of gauge SSB in the SM.
As~noted above,  it has been established quite firmly by now 
that an $\SU(2)$-invariant formulation of the EW model is possible, which decouples the notion of $\SU(2)$ ``breaking" from the effective mass generation mechanism stemming from the EW vacuum transition. 
The very notion of dynamical gauge SSB is thus challenged: 
What is dynamical is the mass generation mechanism, the $\SU(2)$ ``reduction" is kinematical. 

This ``kinematical view" on gauge symmetry reduction is another insight that laid dormant, this time in the mathematical physics literature,  which long prefigured it by noticing in the late 70s and early 80s that ``gauge SSB" mechanisms could be couched in terms of a standard result of fiber bundle geometry, the Bundle Reduction Theorem: 
See e.g. \cite{Trautman, Westenholz, Honda1980, Nikolova-Rizov1984, Bleecker1985} as well as the very nice ending section 5.13 of \cite{Sternberg}.
The theorem is stated as follows:
Given a $G$-principal bundle $P$, and a subgroup $K \subset G$, consider the associated bundle $E$ with typical fiber  the homogeneous space $G/K$.
If there exists a global section $s \in \Gamma(E)$ -- equivalently, a tensorial 0-form $u \in \Omega^0_\text{tens}(P, \ell)$, $\ell$ the left action of $G$ on $G/K$, i.e. $u(pg)=g\-u(p)$\footnote{In case $G/K\subset G$ is a group, such an object also come to be known as a ``dressing field". See e.g. \cite{Francois2018, Francois2023-a}.} -- then the $G$-bundle $P$ as a $K$-principal subbundle $P'\subset P$. 
Actually $G$ acts transitively on the space of $K$-reductions of $P$, which define the same abstract $K$-principal bundle $P'$, so that  it is ``foliated" by such subbundles, $P\simeq P' \times_K G$. 

A section as above arises naturally from 
a tensorial 0-form $\vphi \in \Omega_\text{tens}(P, \rho)$ -- equivalently, from the corresponding associated bundle section  -- 
where the action of $G$ on the representation space $V$ has a cross section whose points have common stability group $K\subset G$. 
One thus has the decomposition $V=\Gamma \times G/K$, with $\Gamma$ the set of $G$-orbits, which implies $\vphi=(\uprho,u)$.
The map $\uprho$ is $G$-invariant (a genuine physical field), while $u$ is the reduction map above. 

In~the case of the SM, $G=U(1) \times S\!U(2) \times S\!U(3)$
and  $V=\CC^2$, so that $\vphi$ is the scalar field providing the $G/K$-valued reduction map,  for $K=U(1)\times S\!U(3)$. Thus, the reduction $P'$ is a $U(1)\times S\!U(3)$-subbundle and one has that $P\simeq P' \times S\!U(2)$. 
In other words, the spacetime bundle has a privileged foliation parametrized by, i.e. is trivial along, the internal space $S^3\sim S\!U(2)$.
This means in particular that the restriction of the fields $P$  to the leaves $P'$ are $\SU(2)$-invariant, i.e. the subgroup $\SU(2) \subset \G=\U(1)\times \SU(2) \times \SU(3)\simeq \Aut_v(P)$ is the Killing group of all fields,  $\SU(2)={}^{\text{gt}}{\sf K}_\upphi$.

The emerging interpretive picture of the SM  could be the following. 
The invariance of the SM physics under the  gauge group $\G$ signals that the gauge d.o.f. of the fields of the SM relationally co-define an enriched spacetime whose points have internal structure $S^1 \times S^3 \times \{{\sf P}^8 \xrightarrow{\text{\tiny $S^3$}} S^5\}$.
Yet, the very \emph{existence} of the (Higgs) scalar field signals that spacetime is trivial along the $S^3$ internal ``direction", which amounts to the statement that the $\SU(2)={}^{\text{gt}}{\sf K}_\upphi \subset \G$ gauge subgroup is a Killing symmetry of all fields. 
This would be the ontological ground behind the notion that $\SU(2)$ is ``broken" in the massive phase of the SM.
We stress that this picture holds as well in the hypothetical massless phase of the SM. 
It is a kinematic, qualitative feature, independent of the dynamics of the scalar field, in particular of the shape of its potential. 
The hypothetical ``spontaneous" change of this shape, i.e. the EW vacuum  transition, dynamically generating masses for fields interacting with the scalar field via Yukawa couplings, is a completely separate issue.

Finally, in echo to section \ref{Killing symmetries}, observe that there is no tension between the claim that gauge physics is gauge invariant and the fact that  non-trivial Killing gauge symmetries are physically meaningful: the latter signal additional physical properties.  
In  the SM case, it signals the existence of the (Higgs) scalar field, i.e. the triviality of the internal structure of spacetime along $S^3$. 
In a theory (a universe) with the same gauge group but witout $\SU(2)$ Killing symmetries, spacetime would not have such trivial internal direction, there would be no (Higgs) scalar field, no notion of fundamental masses as we know it.



\subsubsection{The Aharonov-Bohm effect}
\label{The Aharonov-Bohm effect}

The AB effect was discovered first by Ehrenberg and Siday  in 1949 \cite{Ehrenberg-Siday1949} and again  in 1959 by Aharonov and Bohm \cite{AB1959, AB1961}. See \cite{Hiley2013} for an historical account, \cite{Peshkin-Tonomura1989} for a theoretical and experimental review.
For Aharonov and Bohm, the effect highlights the fundamental importance, and the reality, of the electromagnetic (EM) gauge potential $A$. 
It~is emblematic of gauge theory, and a necessary stop for many philosophers of science interested in the ontology of gauge symmetries \cite{Castellani-Brading2003,Nounou2003, Healey2009, Dougherty2017, Dougherty2021}.

The most often discussed setup (the magnetic AB effect) is that of a modified double slit experiment where a solenoid stands behind the first screen, between the two slits. When  the solenoid is traversed by a current, the interference pattern formed by the electrons on
the second screen is shifted by a phase factor depending only on the flux of the EM field strength $F$ inside the solenoid:
$e^{i\int_c A}=e^{i\int_{\bf s} F}$, where $c$ is a closed path from the source of electrons beam through the two slits to a point on the 
screen, and enclosing the surface $\bf s$ traversed by the solenoid.

Usually, the puzzling aspect of the AB effect is recounted as follows. 
Outside the solenoid, the region accessible
to the electrons described by $\phi$, the EM field strength vanishes, $F=0$, and only the EM potential $A$ is non-zero. So, $A$  is the only local variable that is available to provide an  explanation of the
alteration of the behavior of the electrons via a local interaction between two fields, $A$ and $\phi$.
Yet, $A$ is not gauge-invariant, hence non-physical, contrary to $F$. It seems there is no local gauge-invariant field to account for the alteration of the phase of the electrons along their trajectories.
The natural reaction to this account has often been to conclude that there is a form of non-locality -- or non-separability -- at play, typical of gauge theories and distinct from quantum non-locality, see e.g. \cite{Healey1997, Nounou2003, Lyre2004, Dougherty2017, Francois2018}. 

Of course, one may be puzzled by this account, as it omits that the matter field $\phi$ describing electrons is no more gauge-invariant than the EM potential $A$. So, the lack of local gauge-invariant fields in the AB effect is a general feature of GFT, and an explanation of the effect is but one more instance of the more general problem of extracting observables and interpreting the physics of GFT.

Within the interpretive framework presented above, one may analyse the AB effect in two steps. 
First, one notices that the description in terms of the pair of fields $(A, \phi)$ is  ``bundle-local": these are the local representatives, in a given bundle coordinate system over $P_{|U}$, of the intrinsic object $(\omega, \upphi)$.
Here, $P$ is a $U(1)$-principal bundle, $U\subset M$ represents the region containing the apparatus, 
$\omega$ is a principal connection and $\upphi$ is a (spinor-valued) tensorial 0-form on $P$.
The minimal coupling between matter and the EM field is represented by the tensorial 1-form $D\upphi=d\upphi+\rho_*(\omega) \upphi$.\footnote{The full bundle $P$ is actually an $S\!O(1,3)\times U(1)$-principal bundle, so that the matter field is indeed represented by a tensorial form $\upphi\in \Omega^0_{tens}(P, \rho)$, where $(\rho, S)$ is a spinorial representation for $S\!O(1,3)$, which also happens to be a representation space for $U(1)$. The~connection on $P$ is $\b\omega=\omega+ \omega'$, where $\omega$ is a $U(1)$-connection and  $\omega'$ is a $S\!O(1,3)$-connection. Then $D\upphi$ represents both the EM and gravitational minimal couplings. Of course, in discussing the AB effect, we neglect the coupling to gravity.}
In this intrinsic description on $P$, naturally all objects are bundle-coordinate independent, i.e. invariant under \emph{passive} gauge transformations. 
So, one may say that there is a local gauge-invariant description of the AB effect on $P$: in terms of the interaction between $\omega$ and $\upphi$ -- i.e. the parallel transport of $\upphi$ -- along the horizontal lift $\t c \in P_{|U}$ of the curve $c \in U$. The AB effect is the statement that $\t c$ is not closed whenever $\omega$ is not flat, $\Omega\neq 0$ -- i.e. the holonomy of $\omega$ is non-trivial. 
This account reflects the step \eqref{O1'} in Diag.\ref{Diag2}. 

Now, the final step must take into account the fact that the physics of the AB effect is also invariant under $\Aut_v(P)$, i.e. under \emph{active} gauge transformations. 
Which means that the physical EM and matter fields are described by the $\Aut_v(P)$-class of $(\omega, \upphi) \in \S$. 
What is $\Aut_v(P)$-invariant is the relational (relative) phase of $\omega$ and $\upphi$ along $\t c$. 
More precisely, the physical internal points along the path represented by $\t c$ and the physical values of this relative  phase $\theta$ co-define each other: 
\vspace{-2mm}
\begin{equation}
 \begin{aligned}
 \label{internal-PC-arg-ABeff}
 \theta :  \S \times P\  &\ \rarrow \ \   \S \times P /\sim \  \ \xrightarrow{\simeq} \  \text{Relational internal physical d.o.f.,} \\
 \big(\omega, \upphi; \t c \big) \  &\  \mapsto  \ \big(\omega, \upphi; \t c \big) \sim \big(\psi^*\omega, \psi^*\upphi; \psi\-(\t c)\big)   \ \mapsto\   \theta \big(\omega,\upphi; \t c \big) =\theta \big(\psi^*\omega, \psi^*\upphi; \psi\-(\t c)\big),
 \end{aligned} 
\end{equation}
where $\psi \in \Aut_v(P)$. 
This is an example of the formal expression of the internal point-coincidence argument~\eqref{int-PC-arg},  consistent with the last step \eqref{O2'} in Diag.\ref{Diag2}.
The extension to the general-relativistic description \eqref{O1"}-\eqref{O2"}, Diag.\ref{Diag3}, is obvious.
This relational account of the AB effect is local, in the field theoretical meaning of the term; resulting from the direct interaction of physical fields (d.o.f.) in  arbitrarily small regions (idealised as  points in the limit) of the enriched spacetime, represented by the $\Aut_v(P)$-class of the $U(1)$-bundle $P$. 
We may thus view the AB effect as an instance of non-trivial parallel transport.
In that respect, it is analogous to gravitational non-trivial parallel transport  effects, such as 
gravitationally induced phase shift in photons \cite{Hilweg2017}, or, indeed,  the gravitational AB effect \cite{Hohensee-et-al2012, Overstreet2022}, that we account for in essentially the same way as above. 
In both cases, we see no need to appeal to topological considerations.\footnote{The ``topological" explanations of the AB effect, see e.g. \cite{Nounou2003, Dougherty2017}, rely on the observation that the region accessible to the electron is not simply connected. We observe that this region, $U\backslash \bf s$, is the domain  of the wave-function $\Psi(x)$ of a single electron, not the spacetime region (represented by $U$) pervaded -- co-defined -- by the physical electron field (represented by $\phi$).}



 \section{Conclusion}
\label{Conclusion}

There are a number of other topics that would deserve further discussion within the bundle-geometric-relational interpretive framework of gRGFT defended here. 
Notably, the issues of coupling constants and Noether theorems. 
In our view, the latter, as it relates  symmetries to conserved charges, should be seen as essentially  just a heuristics: 
It is much more enlightening to understand conserved charges as  a direct consequence of the dynamics, i.e. as derived entirely from the field equations. Indeed it is where the physical explanatory power resides.
The ``magic" of deriving conserved current from symmetry considerations arises, in gRGFT, from the fact that the gauge potential is Lie algebra-valued. 
About coupling constants (featuring in covariant derivatives), we may say that they essentially measure the sensibility, and strength of the response, of a field to the geometry of the relevant part of spacetime internal structure.
The fact that not all fields have non-zero coupling would reflect their inability to probe part of this internal structure, 
while the fact that the coupling is not the same for all fields excludes a ``generalised equivalence principle" for all interactions. 
We will flesh out these terse 
remarks somewhere else.
\smallskip

The standard formulation of gauge theory requires to work on $M$, the base of the principal bundle $P$ (and of its associated bundles). 
It means that, to all practical purposes, one works with local representatives on $U\subset M$ of intrinsic objects of $P$, i.e. within a bundle coordinate system. 
Aside from a few examples (e.g. \cite{Neeman-Regge1978b,Ross:1996aa,Helein:2017,Castellani:2018zey,Castellani:2019pvh,Castellani:2022iib,Helein:2023qpt}) there have not been  focused and intentional efforts to formulate physics directly on a bundle space. 
As a result, we do not have a mature formalism to express physics in an bundle-intrinsic way. 
A situation which is surprising if one takes seriously, as we have argued one should, the view that gauge theory signals that the geometry of the physical enriched spacetime is that of a principal bundle, as expressed in the second box of Diag.\ref{Diag3} and \eqref{O1"}.
The situation may be considered as unsatisfactory as the one we might have found ourselves in if, after  understanding general-relativistic physics and realising that the geometry of physical spacetime is that of a manifold $M$, as expressed in the second box of  Diag.\ref{Diag1} and \eqref{O1}, Physics never got past using coordinate representatives, i.e. continued working up-to coordinate changes, and never developing (or using) intrinsic, coordinate-independent, differential geometric methods to formulate GR physics on $M$.
We know, as a matter of fact, the technical usefulness and conceptual payoff of such intrinsic methods. So we easily understand the missed opportunities if the above had come to pass. 

We are of the opinion that there should be a more concerted effort to formulate physics intrinsically on $P$ -- i.e. in a ``passively gauge invariant" way. 
Most of the key objects (fields) are already well-defined there. 
Usually one works locally because a metric field can be defined on $U\subset M$, which also induces a Hodge star operator. This allows two things at once: 
First, Hodge duality allows to write Lagrangian functionals in an intrinsic (coordinate-independent) way on $U\subset M$, secondly  it represents the (non-minimal) coupling to gravity.\footnote{For spinors, this coupling is minimal, via the covariant derivative, i.e. the twisted Dirac operator.}
To work on $P$, one must fulfill these two desiderata: either defining an equivariant metric on $P$ from which to derive a notion of Hodge duality, or find a replacement to the latter and implement coupling to gravity separately -- e.g. through the soldering form, a.k.a. the tetrad field, defined on $P$ already.
The availability of an intrinsic formulation of gRGFT on $P$ would probably make it easier to accept the ontology of the bundle as representing a physical enriched spacetime -- if only for psychological reasons. 
The problem is well circumscribed and does not seem unsurmountable. 

\smallskip

But accepting this ontology is but the first of two steps as illustrated in Diag.\ref{Diag3}, whereby the conjunction of the generalised hole argument and point-coincidence argument leads to the relational picture \eqref{O1"b}. 
In the standard mathematical formulation of gRGFT, relationality of physics is tacit, encoded in the manifest invariance, or covariance, of the theory under  $\Diff(M)$ and $\Aut_v(P)\simeq \H$ (i.e.  $\Aut(P)$). 
It can thus be easily overlooked, which may lead to a number of unfortunate misconceptions. 
For instance, the notion that spacetime ``boundaries" (at finite distance) break $\Diff(M)$ or $\H$-gauge symmetries, which has the same conceptual structure as a hole argument. 
This notion evaporates once it is recognised that a \emph{physical} boundary is relationally defined, and is invariant (under $\Diff(M)$, $\Aut_v(P)\simeq \H$, or $\Aut(P)$).\footnote{A popular strategy to resolve such ``boundary problem" is to add extra d.o.f. at the boundary (sometimes called ``edge modes"). This,~ultimately, should be understood as an attempt to restore relationality. The more common strategy of imposing boundary conditions does nothing to resolve the conceptual physical  issue.}
This and similar misconceptions, together with the various  countermeasures put forward to solve the perceived issue, would be avoided had one a framework in which both relationality and strict invariance are manifest.
It is so because such a framework would have the significant advantage of giving immediate access to physical relational observables.
We will put forward a proposal for such a manifestly relational framework in a separate work 
\cite{JTF-Ravera2024}.

\section*{Acknowledgment}

J.F. is supported by the OP J.A.C. MSCA grant, number CZ.02.01.01/00/22\_010/0003229, co-funded by the Czech government Ministry of Education, Youth \& Sports and the EU,
as well as  the Austrian Science Fund (FWF), \mbox{[P 36542]}. 
L.R. would like to thank the DISAT of the Polytechnic of Turin and the INFN for financial support.

{
\small
 \bibliography{philobib}
}

\end{document}